\documentclass[fleqn,usenatbib]{mnras}

\usepackage{newtxtext,newtxmath}
\usepackage{caption}
\usepackage{subcaption}
\usepackage{color}
\usepackage{xcolor}

\usepackage[T1]{fontenc}

\DeclareRobustCommand{\VAN}[3]{#2}
\let\VANthebibliography\thebibliography
\def\thebibliography{\DeclareRobustCommand{\VAN}[3]{##3}\VANthebibliography}

\usepackage{graphicx}	
\usepackage{amsmath}	

\title[Double pulsar eclipse polarimetry]{MeerKAT observations of pair-plasma induced birefringence in the double pulsar eclipses}

\author[M. E. Lower et al.]{\parbox{\textwidth}
{
M.~E. Lower,$^{1}$\thanks{E-mail: {\href{mailto:marcus.e.lower@gmail.com}{marcus.e.lower@gmail.com}}}
M. Kramer,$^{2,3}$
S. Johnston,$^{1}$
R.~P. Breton,$^{3}$
N. Wex,$^{2}$
M. Bailes,$^{4,5}$
S. Buchner,$^{6}$
F. Camilo,$^{6}$\\
L.~S. Oswald,$^{7,8}$
D.~J. Reardon,$^{4,5}$
R.~M. Shannon,$^{4,5}$
M. Serylak,$^{9,10}$
and V. Venkatraman~Krishnan$^{2}$
}
\\ \\
$^{1}$Australia Telescope National Facility, CSIRO, Space and Astronomy, PO Box 76, Epping, NSW 1710, Australia\\
$^{2}$Max-Planck-Institut f\"{u}r Radioastronomie, Auf dem H\"{u}gel 69, D-53121 Bonn, Germany\\
$^{3}$Jodrell Bank Centre for Astrophysics, School of Physics and Astronomy, The University of Manchester, M13 9PL, UK\\
$^{4}$Centre for Astrophysics and Supercomputing, Swinburne University of Technology, PO Box 218, Hawthorn, VIC 3122, Australia\\
$^{5}$Australia Research Council Centre for Excellence for Gravitational Wave Discovery (OzGrav)\\
$^{6}$South African Radio Astronomy Observatory, 2 Fir Street, Black River Park, Observatory 7925, South Africa\\
$^{7}$Department of Astrophysics, University of Oxford, Denys Wilkinson Building, Keble Road, Oxford OX1 3RH, UK\\
$^{8}$Magdalen College, University of Oxford, Oxford OX1 4AU, UK\\
$^{9}$SKA Observatory, Jodrell Bank, Lower Withington, Macclesfield SK11 9FT, UK\\
$^{10}$Department of Physics and Astronomy, University of the Western Cape, Bellville, Cape Town 7535, South Africa
}

\date{Accepted XXXX. Received YYYY; in original form ZZZZ}

\begin{document}
\label{firstpage}
\pagerange{\pageref{firstpage}--\pageref{lastpage}}
\maketitle

\begin{abstract}
PSR~J0737$-$3039A/B is unique among double neutron star systems.
Its near-perfect edge-on orbit causes the fast spinning pulsar A to be eclipsed by the magnetic field of the slow spinning pulsar B.
Using high-sensitivity MeerKAT radio observations combined with updated constraints on the system geometry, we studied the impact of these eclipses on the incident polarization properties of pulsar A. 
Averaging light curves together after correcting for the rotation of pulsar B revealed enormous amounts of circular polarization and rapid changes in the linear polarization position angle, which occur at phases where emission from pulsar A is partially transmitted through the magnetosphere of pulsar B.
These behaviours confirm that the eclipse mechanism is the result of synchrotron absorption in a relativistic pair-plasma confined to the closed-field region of pulsar B's truncated dipolar magnetic field. 
We demonstrate that changes in circular polarization handedness throughout the eclipses are directly tied to the average line of sight magnetic field direction of pulsar B, from which we unambiguously determine the complete magnetic and viewing geometry of the pulsar.
\end{abstract}



\section{Introduction}

The double pulsar system PSR~J0737$-$3039A/B is an extraordinary astrophysical laboratory that offers us a chance to directly probe the magnetic field of an active pulsar.
Due to its almost edge-on orbit ($i = 90.63^{\circ} \pm 0.03^{\circ}$; \citealt{Hu2022, Askew2024}) the 22.7\,ms PSR~J0737$-$3039A (`pulsar A'; \citealt{Burgay2003, Kramer2021b}) is eclipsed for approximately 30-40\,s every 2.4\,hrs by the magnetosphere of the slower 2.77\,s PSR~J0737$-$3039B (`pulsar B'; \citealt{Lyne2004}).
Due to the intense particle wind from pulsar A, the size of the eclipse region is less than a third the nominal light-cylinder radius of pulsar B's dipolar magnetic field, indicating much of its magnetosphere is blown backwards into a magnetotail \citep{Lyutikov2004, Arons2005}.
High-time resolution observations revealed the eclipse light curves exhibit peaks and troughs that repeat at once and twice the 2.8\,s spin-period of pulsar B \citep{McLaughlin2004}.
This behaviour was suggested by \citet{Lyutikov2005b} to be the result of radiation from pulsar A undergoing synchrotron absorption in a relativistic electron-positron pair plasma trapped within the closed-field region of pulsar B's magnetosphere.
The appearance of the `transparency windows' at different eclipse phases, where our line of sight to pulsar A is unobstructed by the magnetic field of pulsar B, is simply a result of the geometric orientation of the pulsar spin and magnetic axes.
This simple model has been remarkably successful at replicating the observed eclipse light curve, and has enabled unique tests of relativistic spin-orbit coupling and geodetic precession in strong gravity \citep{Breton2008, Lower2024}.

In addition to enabling geometric measurements, the eclipses also provide a novel means to directly probe the plasma surrounding pulsar B through studying changes in the polarimetry of pulsar A over time and observing frequency. 
Previous attempts to search for such spectropolarimetric variability have only been marginally successful owing to the limited instantaneous sensitivity of previous generation radio telescopes.
\citet{Yuen2012} found 2-$\sigma$ evidence for a change in the linear polarization position angle of pulsar A within the average eclipse envelope that depends on observing frequency.
This effect was attributed either to Faraday rotation induced by a population of mildly relativistic charged particles in the magnetotail of pulsar B altering the rotation measure (RM) of pulsar A, or preferential absorption of one linear polarization over the other.
Their limited time resolution meant they were insensitive to rapid changes in the overall polarization content throughout the eclipses.
Improved polarimetric measurements during the eclipses require collecting high signal-to-noise ratio (S/N) spectropolarimetric observations, which can be achieved through observing the double pulsar with more sensitive radio telescopes, coherent averaging of multiple eclipses together after correcting for the changing rotation phase of pulsar B, or both.

Following on from our previous inference of the geometric orientation and precession of pulsar B from modelling MeerKAT total intensity light curves \citep{Lower2024}, we have produced high S/N rotation-phase corrected eclipse polarization light curves.
This has enabled us to conduct a full spectropolarimetric analysis of the double pulsar eclipses, the results of which we present in this work.
Our observations with MeerKAT and data processing steps are outlined in Section~\ref{sec:obs}.
The data analysis techniques we employed and results are detailed in Section~\ref{sec:analysis}.
We discuss the interpretation of these results in Section~\ref{sec:disc}.
Finally, we summarise our findings and make concluding statements on future polarimetric studies of the double pulsar in Section~\ref{sec:conc}.

\section{Observations} \label{sec:obs}

\begin{figure}
    \centering
    \includegraphics[width=\linewidth]{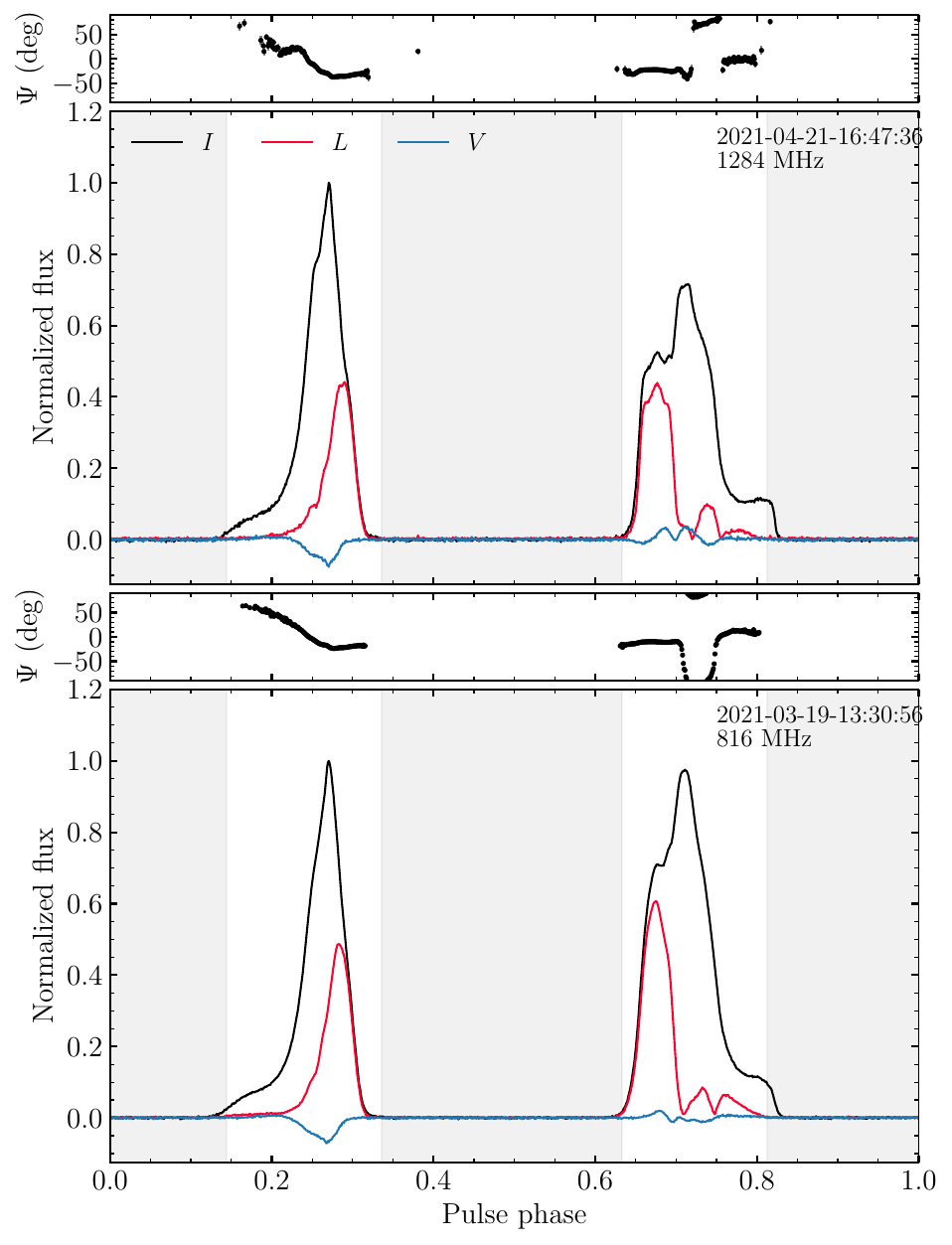}
    \caption{Example time- and frequency-averaged pulsar A polarization profiles detected by MeerKAT at L-band (top) and UHF (bottom). The upper panels show the linear polarization position angle ($\Psi$), while the lower panels depict the total intensity (black), linear polarization (red) and circular polarization (blue) profiles. The grey-regions indicate the off-pulse regions that were excluded when extracting the polarization light curves.}
    \label{fig:profile}
\end{figure}

We observed the double pulsar with an approximately monthly cadence using the L-band (1248\,MHz) and UHF (816\,MHz) receiver systems of the 64 antenna SARAO Meer Karoo Array Telescope (MeerKAT; \citealt{Jonas2016}) under the `relativistic binary' programme of the MeerTime large survey project \citep{Bailes2020, Kramer2021a}.
We note that the recorded data were subsequently averaged in frequency from 1024 to 512 channels to reduce data storage needs. 
Our data processing approach was previously described in \citet{Lower2024} and a detailed breakdown of the data recording process, folding and cleaning steps can be found therein.
One key difference to the previous analysis is that we retain both frequency and polarization information for this work.
In \citet{Lower2024} only the frequency-averaged total intensity data were analysed.
Our polarization calibration procedure followed the process outlined in \citet{Serylak2021}, after which we corrected the two linear polarizations (Stokes $Q$ and $U$) for Faraday rotation in the interstellar medium using the ${\rm RM} = 120.82$\,rad\,m$^{-2}$ measured by \citet{Kramer2021a}.
Throughout this work we refer to the handedness of the circular polarization in the IEEE/pulsar convention \citep{vanStraten2010a}, where positive Stokes $V$ indicates left-hand circular (LHC) and negative Stokes $V$ is right-hand circular (RHC).
Figure~\ref{fig:profile} shows the time- and frequency-averaged polarization profile of pulsar A detected at both L-band and UHF.
Our eclipse light curves were generated by averaging together the polarization data contained within two `on-pulse' windows (non-grey regions) that covered the main- and inter-pulse profile components. 

\section{Analysis and Results}\label{sec:analysis}

\subsection{Incoherent averaging and Faraday rotation variations}

\begin{figure}
    \centering
    \includegraphics[width=\linewidth]{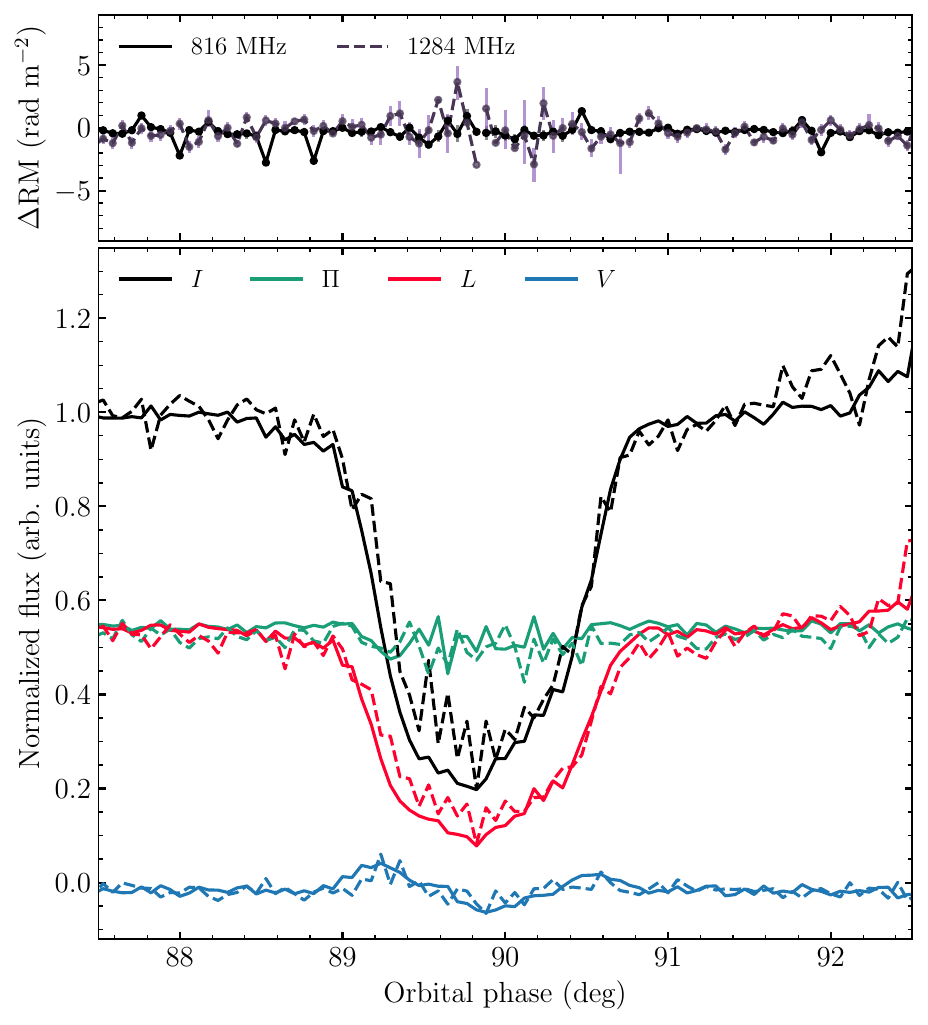}
    \caption{Polarimetric analysis of incoherently averaged UHF (solid lines) and L-band (dashed lines) eclipses. The upper panel shows the differential RM of pulsar A measured in the two bands. The lower panel depicts the rotation-phase and frequency averaged polarization light curves, where total intensity is shown in black, polarization fraction in green, total linear polarization in red, and circular polarization in blue.}
    \label{fig:incoherent}
\end{figure}

To enhance our sensitivity to changes in the bulk polarization properties of pulsar A, we averaged together 16 L-band and 33 UHF eclipses collected between 2020 February and 2022 August in orbital phase.
This was done without accounting for the differences in pulsar B rotation phase at each epoch, producing the two high S/N `incoherent' light-curves shown in Figure~\ref{fig:incoherent}. 
The incoherently averaged light-curves at L-band and UHF appear qualitatively similar in total intensity as well as linear and circular polarization.
Both bands display an enhancement in the detected circular polarization throughout the eclipse region. 
This appears as excess LHC in the ingress and egress phases, and RHC around superior conjunction.
The total polarization fraction ($\Pi = \sqrt{Q^{2} + U^{2} + V^{2}}/I = P/I$) remains largely constant throughout the entire eclipse envelope, indicating the excess circular polarization at these phases is compensated by a drop in linear.
This constant $\Pi$ differs from the expected increase of $\Delta\Pi \sim 0.3$ at the deepest part of the eclipse \citep{Lyutikov2005b}, though it is worth noting that these predictions were for a coarsely sampled single eclipse of an unpolarized background object. 
By contrast, the light curves in Figure~\ref{fig:incoherent} were made from averaging real observations of pulsar A, which displays a relatively high linear polarization fraction.

We searched for signs of Faraday rotation induced by non-relativistic charged particles in the magnetosphere and magnetotail of pulsar B by performing direct fits to Stokes $Q$ and $U$ spectra extracted from each orbital phase bin using {\sc RMNest} \citep{Lower2022}.
Uniform priors were assumed between $-1000\,{\rm rad\,m^{-2}} < \pi({\rm RM}) < 1000\,{\rm rad\,m^{-2}}$ for the RM and $-90^{\circ} < \pi(\Psi_{0}) < 90^{\circ}$ for the reference position angle.
We sampled the posterior distributions for the model parameters using a Gaussian likelihood and the {\sc PyMultiNest} nested sampling algorithm \citep{Buchner2014}.
The resulting differential RM values are displayed in the upper panel of Figure~\ref{fig:incoherent}.
Neither the L-band nor UHF measurements display a substantial deviation away from $\Delta{\rm RM} = 0$\,rad\,m$^{-2}$ within the eclipse envelope or immediately following the egress phase.
The 1-$\sigma$ scatter in the median recovered RM of $\sigma_{\rm RM, L-band} = 1$\,rad\,m$^{-2}$ at L-band and $\sigma_{\rm RM, UHF} = 0.6$\,rad\,m$^{-2}$ at UHF.
While there is excess scatter in the L-band measurements towards superior conjunction, this likely arises from the comparatively lower signal-to-noise ratio of the L-band data and there being fewer available eclipses to average together.
Hence, it is unlikely that excess RM imparted within the magnetotail of pulsar B can explain previous reports of PA swing variations throughout the eclipse envelope \citep{Yuen2014}.

\subsection{Coherently averaged eclipses}

In order to better understand the short-term changes induced in the polarization properties of pulsar A, we also explored an alternative means of averaging the eclipse data together.
This approach involved pre-correcting for the rotation of pulsar B by computing its rotation phase at each epoch from the geometric fits in \citet{Lower2024} and shifting several light curves in time such that they all overlap at the same initial spin phase.
In doing this, we gain sensitivity to polarimetric variations caused by the radio waves from pulsar A cutting through specific regions of pulsar B's truncated dipolar magnetic field that were otherwise averaged out by the incoherent method.
However, only a handful of eclipses could be averaged together at any one time using this approach, as geodetic precession alters both the overall modulation pattern and location of the rotation-phase zero-point over long time scales.

\begin{figure*}
     \centering
     \begin{subfigure}[b]{0.495\textwidth}
         \centering
         \includegraphics[width=\textwidth]{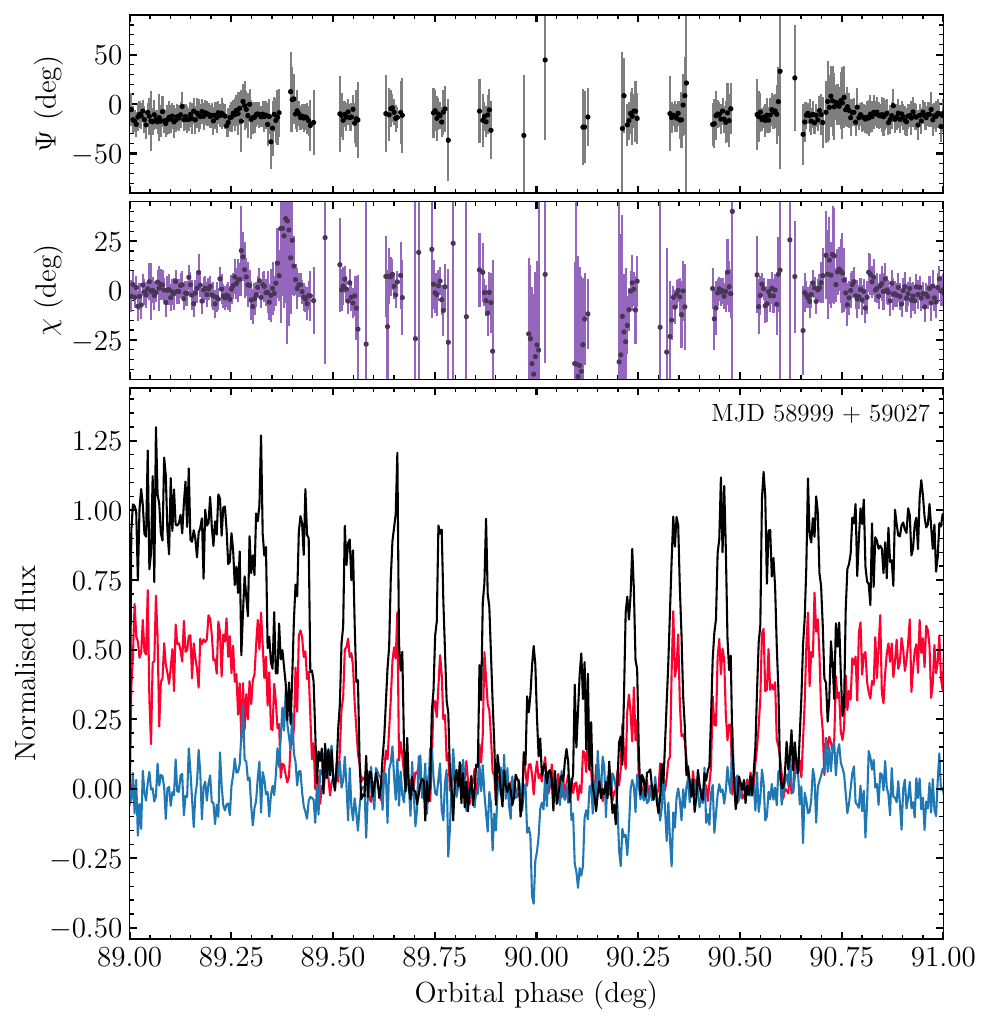}
     \end{subfigure}
     \begin{subfigure}[b]{0.495\textwidth}
         \centering
         \includegraphics[width=\textwidth]{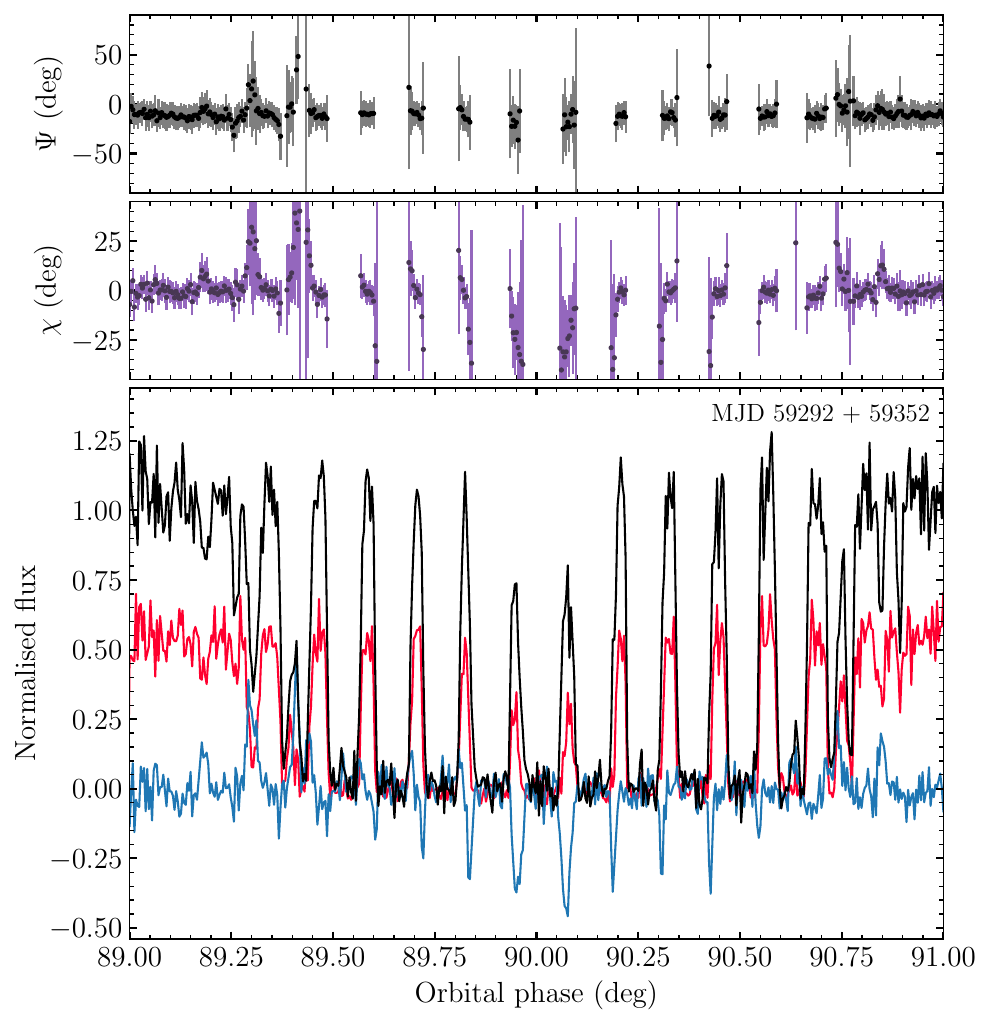}
     \end{subfigure}
     \begin{subfigure}[b]{0.495\textwidth}
         \centering
         \includegraphics[width=\textwidth]{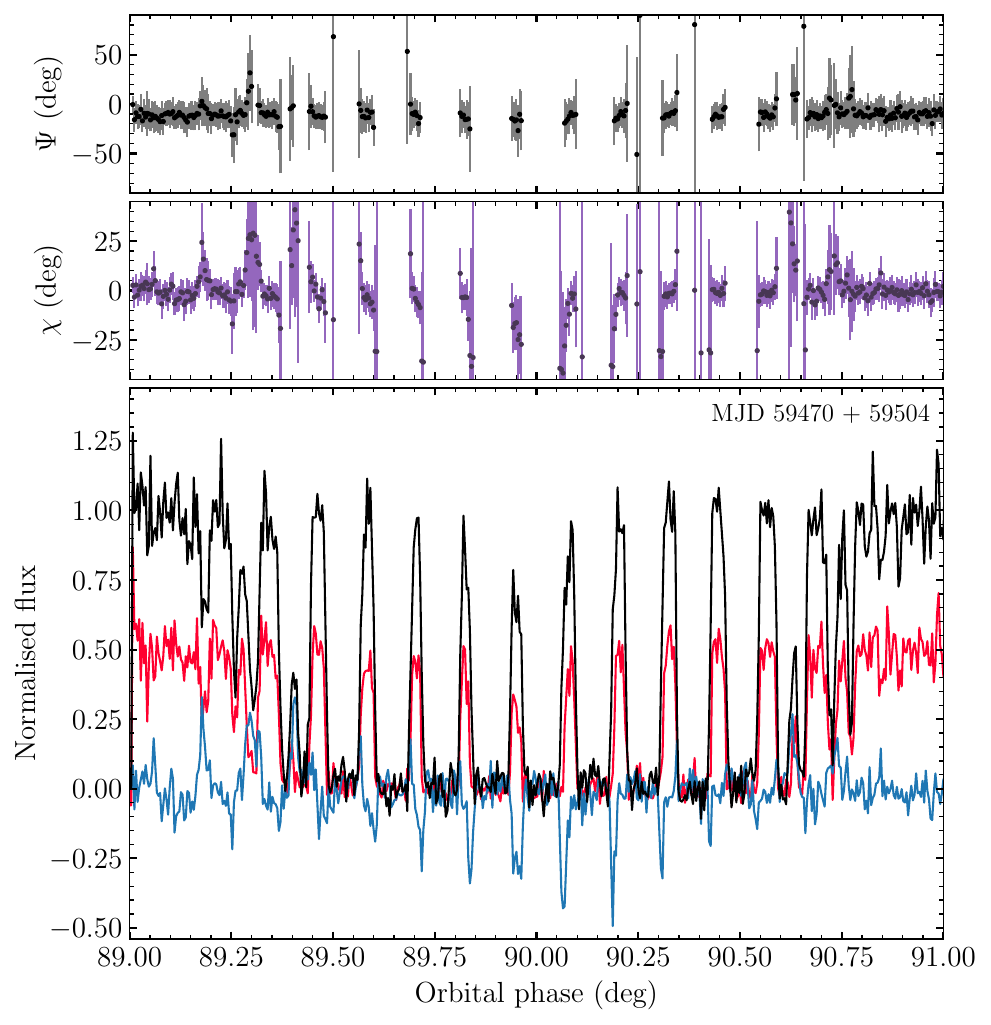}
     \end{subfigure}
     \begin{subfigure}[b]{0.495\textwidth}
         \centering
         \includegraphics[width=\textwidth]{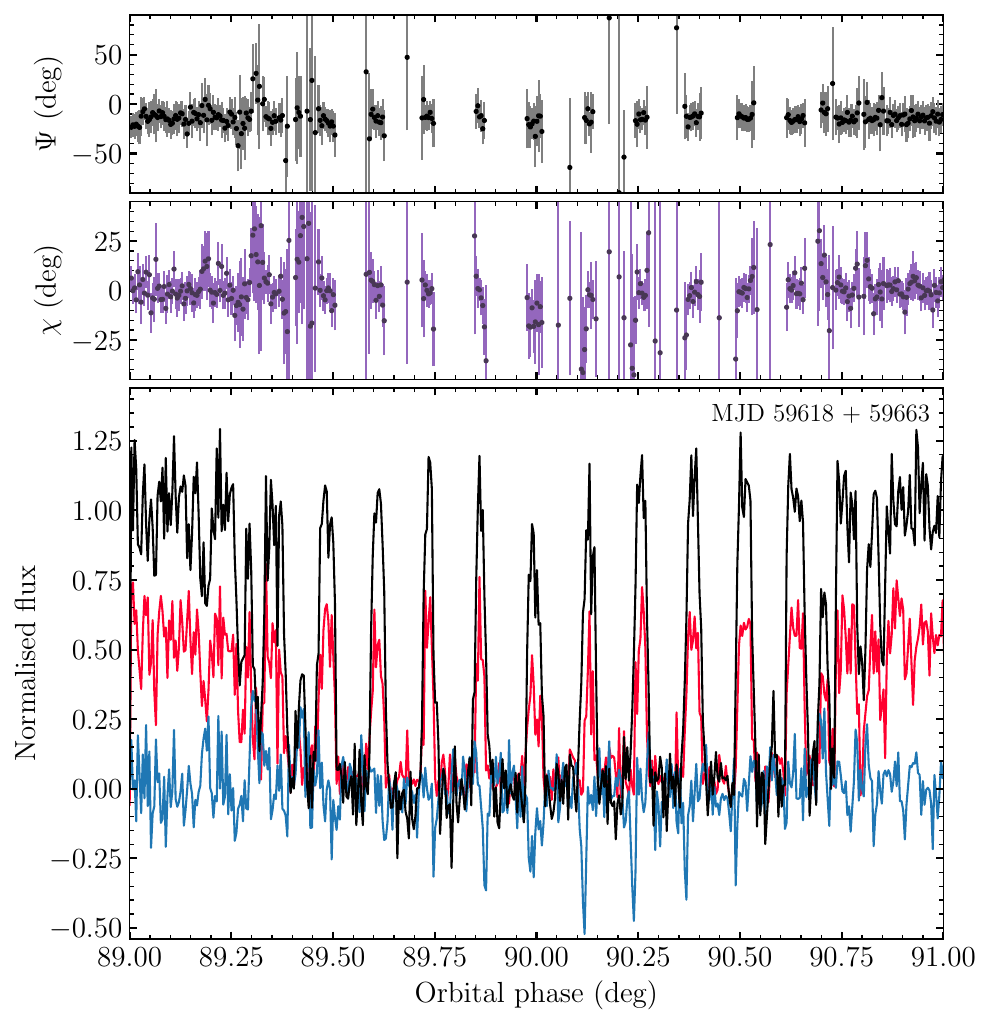}
     \end{subfigure}
        \caption{Examples of UHF eclipse light curves coherently averaged together after correcting for the rotation phase of pulsar B. The top panel of each plot shows the linear polarization position angle, upper-middle panel the ellipticity angle and the bottom panels show the polarization light curve of pulsar A.}
        \label{fig:coherent}
\end{figure*}

Coherently averaging together smaller sub-sets of four near-in-time UHF eclipses (i.e detected within approximately 2 months of one another) after correcting for pulsar B's rotation phase reveals remarkable variability in the polarization state of pulsar A.
Examples of such light curves are presented in Figure~\ref{fig:coherent}. 
As anticipated from our analysis of the incoherently summed light curves, the emergence of intense circular polarization occurs predominately in the `partial transparency' windows seen in the ingress, egress and maximum phases of the eclipse.
Here our line of sight to pulsar A intersects the outermost regions of pulsar B's closed-field lines where the optical depth of pulsar B's magnetosphere is sufficiently low that the radiation from pulsar A is only partially absorbed by the confined plasma.
This is also true for the edges of the `full' transparency windows, where in the early part of the eclipse we detect a spike of LHC when pulsar A enters a transparent window, followed by a sign flip to RHC when it exits the window and is again eclipsed by pulsar B.
The opposite behaviour (RHC then LHC) is seen in the fully transparent windows located in the latter half of the eclipse light curves.
Since the modulation pattern of the eclipse is not being incoherently averaged over, the overall circular polarization fraction is much higher than that depicted in Figure~\ref{fig:incoherent}.
Alongside the enhanced circular polarization, we also detect clear variations in the average linear polarization position angle (PA; $\Psi = 0.5 \tan^{-1}(U/Q)$) of pulsar A during the partial-transparency windows located in the ingress and egress phases.
Marginal differences in PA away from the nominal off-eclipse value are also visible towards the edges of some full-transparency windows.
Significant variations are also seen in the ellipticity angle ($\chi = 0.5 \tan^{-1}(V/L)$) which track the Stokes $V$ enhancements.

\begin{figure}
    \centering
    \includegraphics[width=\linewidth]{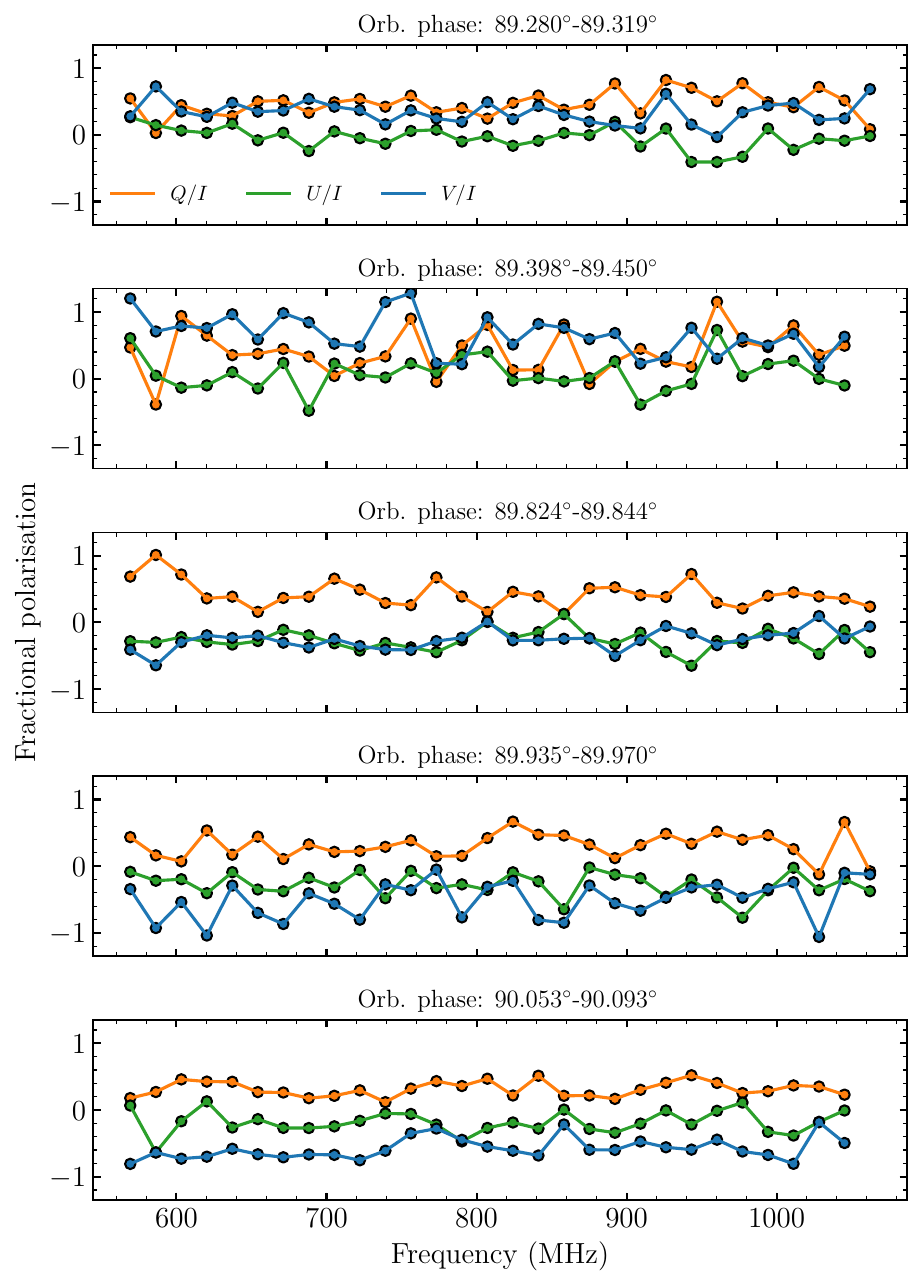}
    \caption{polarization fraction of pulsar A ($Q/I$ in orange, $U/I$ in green and $V/I$ in blue) as a function of frequency extracted from five different phases of the light curve generated from data collected on MJD 59292 and 59352. Apparent over-polarization in the second panel is an artefact of reduced S/N.}
    \label{fig:spectra}
\end{figure}

Birefringent effects such as Faraday conversion\footnote{Sometimes referred to as generalised Faraday rotation} result in frequency-dependent variations between Stokes $Q$, $U$ and $V$ that could explain the large amount of induced circular polarization in Figures~\ref{fig:incoherent} and \ref{fig:coherent}.
The exact frequency dependence induced by Faraday conversion can vary between $\nu^{-1}$ to $\nu^{-3}$ and is set by the physics of the propagating medium \citep[see, e.g.][]{Kennett1998, Gruzinov2019, Lyutikov2022b}.
In a pair plasma, one may expect the electrons and positrons to contribute Faraday conversion with equal and opposite signs, thereby cancelling it out.
However, slight asymmetries in the ratio of one particle over the other can result in a significant amount of detectable circular polarization being produced \citep{Sazonov1969, Noerdlinger1978}.
To test whether this effect can explain the observed increase in circular polarization detected within the eclipses, we searched for clear signs of correlated linear and circular variability within five different regions of the polarization data collected on MJD 59292 and 59352.
The eclipses detected on these dates had particularly high S/N, likely due to favourable refractive scintillation boosting the flux density of pulsar A.
In Figure~\ref{fig:spectra}, we show the resulting polarization fraction of pulsar A in each of these windows as a function of frequency.
The first two panels were taken from partially-transparent windows in the ingress phase, the middle panel was extracted from the edge of a fully transparent window between ingress and superior conjunction, while the final two panels correspond to the two partial-transparency windows detected either side of superior conjunction.
In all five cases, both the linear and circular polarization fractions remain largely constant with frequency with no changes in sign or handedness.
Spectropolarimetric fits using the phenomenological Faraday conversion model in {\sc RMNest} (see \citealt{Lower2021} for details) failed to reconstruct the data, recovering unconstrained posterior distributions for most model parameters. 

\begin{figure}
    \centering
    \includegraphics[width=\linewidth]{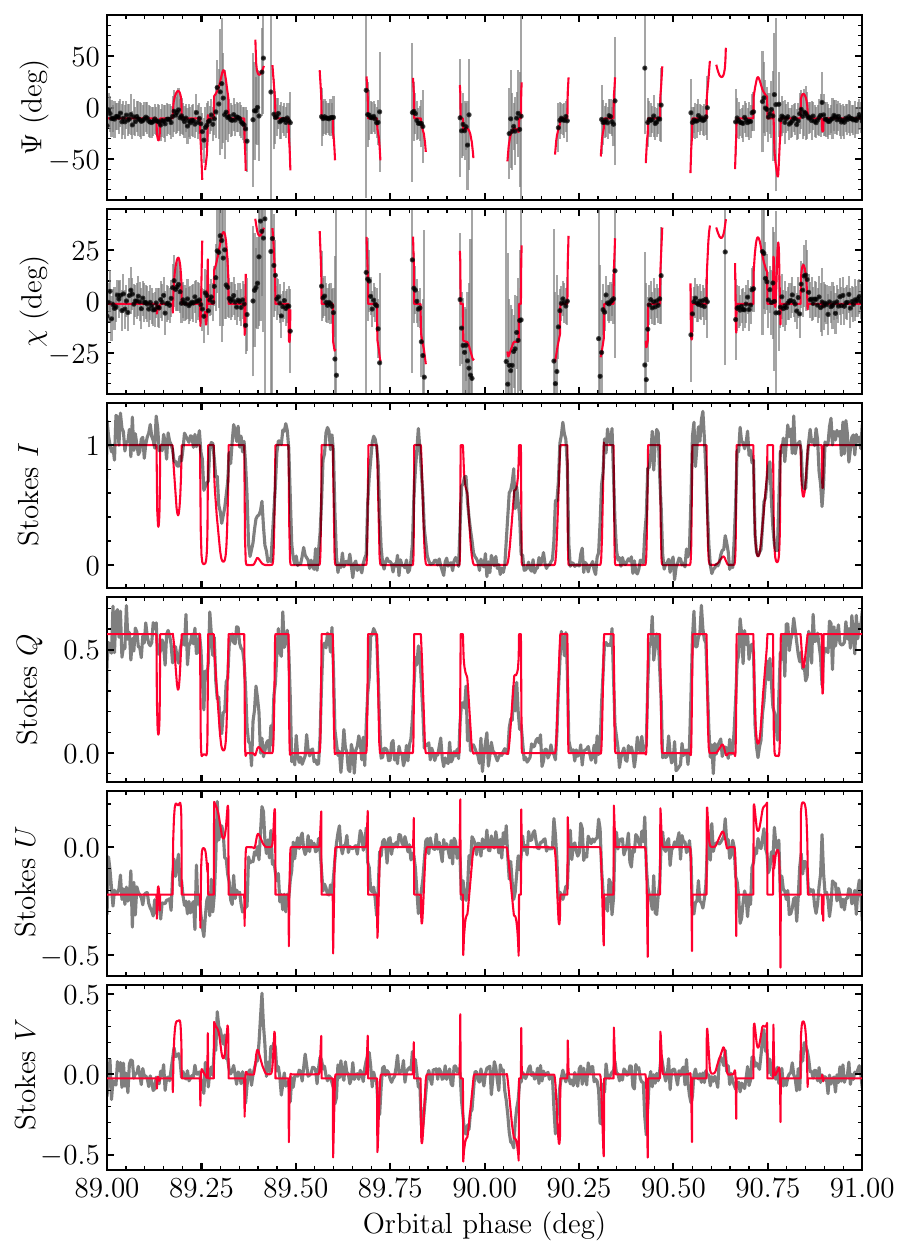}
    \caption{Comparison between the simulated (red lines) and observed (black points or grey lines) polarization properties of pulsar A throughout the combined eclipse light curves detected on MJD 59292 and 59352.}
    \label{fig:comparison}
\end{figure}

In addition to the total intensity model used in \citet{Breton2008} and \citet{Lower2024} for tracking the geodetic precession of pulsar B, \citet{Lyutikov2005b} devised a set of radiative transfer equations that describe the propagation of polarized radiation through the eclipses (see Equations \ref{eqn:stokesIa} to \ref{eqn:stokesV} below).
Note, that equation 64 in \citet{Lyutikov2005b} re-expresses Stokes $I$ and $U$ as a combination of two reference polarization states, $I^{a} = 0.5 (I + Q)$ and $I^{b} = 0.5 (I - Q)$, corresponding to the polarization intensity orthogonal to the orbital plane of the system ($I^{a}$) and parallel to the orbital plane ($I^{b}$).
However, solving these equations as written does not provide a set of polarization light curves.
Instead our numerical solution to them requires making several simplifying assumptions.
First that the angle between the projected magnetic field normal to the orbital plane and our line of sight remains fixed at the incident value ($\chi_{B}(x) = \chi_{B,0}$) at each step through the magnetosphere of pulsar B.
This assumption is largely valid as propagation of radiation through the magnetosphere of pulsar B is significantly faster than changes in magnetic field strength and direction as the pulsar rotates.
And second, that polarization parameters correspond to the unaltered incident values that are intrinsic to pulsar A (i.e $I^{a}_{0}$, $I^{b}_{0}$, $U_{0}$ and $V_{0}$).
With these simplifications in mind, the light-curves for the two reference polarization states along with Stokes $U$ and $V$ were then computed as
\begin{equation}\label{eqn:stokesIa}
\begin{split}
    I^{a} =\, & I^{a}_{0} \Big[ \sin^{4}(\chi_{B,0})\exp(-4\tau_{\nu}/3) + \cos^{4}(\chi_{B,0})\exp(-2\tau_{\nu}/3) \\
    & - \frac{1}{2}\sin^{2}(2\chi_{B,0})\exp(-\tau_{\nu}) \Big] -\frac{1}{4}U_{0}\exp(-2\tau_{\nu}/3)\\ 
    & + \frac{1}{2}V_{0}\sin{2\chi_{B,0}}\exp(-0.54\tau_{\nu}),
\end{split}
\end{equation}
\begin{equation}\label{eqn:polcurve}
\begin{split}
    I^{b} =\, & I^{a}_{0} \Big[ \cos^{4}(\chi_{B,0})\exp(-4\tau_{\nu}/3) + \sin^{4}(\chi_{B,0})\exp(-2\tau_{\nu}/3) \\
    & - \frac{1}{2}\sin^{2}(2\chi_{B,0})\exp(-\tau_{\nu}) \Big] -\frac{1}{4}U_{0}\exp(-2\tau_{\nu}/3)\\ 
    & - \frac{1}{2}V_{0}\sin{2\chi_{B,0}}\exp(-0.54\tau_{\nu}),
\end{split}
\end{equation}
\begin{equation}
\begin{split}
    U =\, & U_{0}\exp(-\tau_{\nu}) - (I^{a}_{0} + I^{b}_{0})\sin(2\chi_{B,0})\exp(-2\tau_{\nu}/3) \\
    & - V\cos(2\chi_{B,0})\exp(-0.54\tau_{\nu}),
\end{split}
\end{equation}
\begin{equation}\label{eqn:stokesV}
\begin{split}
    V = & V_{0}\exp(-\tau_{\nu}) - [(I^{a}_{0} - I^{b}_{0})\sin(2\chi_{B,0})\\
    & - U_{0}\cos(2\chi_{B,0})]\exp(-0.54\tau_{\nu}),
\end{split}
\end{equation}
where $\tau_{\nu}$ is the optical depth given by equations 50 and 52 of \citet{Lyutikov2005b}
\begin{equation}\label{eqn:tau}
    \tau_{\nu} = \int \eta_{\nu} dx = \frac{\mu}{\nu_{\rm GHz}^{\ell}} \int_{-R_{\rm mag}}^{R_{\rm mag}} \Big(\frac{B\sin(\kappa)}{B_{\rm mag}}\Big) d\Big(\frac{x}{R_{\rm mag}}\Big).
\end{equation}
Here, $\eta$ is the synchrotron absorption coefficient, $\mu$ is a parameter that encodes all of the physical properties of the magnetospheric plasma, $\nu_{\rm GHz}$ the radio frequency normalised to 1\,GHz with frequency scaling $\ell$, $R_{\rm mag}$ the truncated radius of pulsar B's magnetosphere, $B$ is the pulsar magnetic field strength in units of $B_{\rm mag}$ (magnetic field strength at $R_{\rm mag}$), and $\kappa$ is the angle between the local magnetic field direction and our line of sight.
The frequency-scaling parameter was previously determined to be consistent with $\ell = 1/3$ from fits to sub-banded MeerKAT UHF and L-band data \citep{Lower2024}.
Using the best-fit pulsar B geometry from \citet{Lower2024} and setting the incident polarization to the median pre-eclipse values of pulsar A, we compare the resulting position and ellipticity angles and Stokes parameters against the polarization light curve obtained from the MJD 59292 and 59352 eclipses in Figure~\ref{fig:comparison}. 
Qualitatively, the simulated polarization properties display a remarkable level of similarity to the observed data.
We recover the same change in circular handedness from LHC to RHC going from ingress to superior conjunction, and then back to LHC during the egress phase during the partial transparency windows.
The same changing handedness behaviour is also seen at the edges of the full-transparency windows.
Similarly, the observed changes in position angle and ellipticity angle are almost fully replicated by the model to within the respective uncertainties.
There are however some notable differences, namely the theoretical model tends to over-predict the linear polarization amplitude within several of the partial-transparency windows.
It also fails to adequately recover the partial-transparency windows at orbital phases $\sim$89.4 and $\sim$90.6 degrees.
However, it has been established that the \citet{Lyutikov2005b} model is less reliable within the ingress and egress phases of the eclipse, where deviations away from the assumed perfect axisymmetric dipole of pulsar B and local distortions in the magnetic field topology may be present \citep{Breton2008, Lower2024}.
Such effects are expected from numerical simulations of the particle wind from pulsar A impinging the magnetosphere of pulsar B (e.g., \citealt{Arons2005, Zhong2024}) and from averaging over small changes in the pulsar geometry as it precesses.
Additionally, the eclipses display substantial epoch-to-epoch variations in morphology and radial extent due to stochastic variations in the confined plasma, which may also affect the observed light curves.

\section{Discussion}\label{sec:disc}

Our MeerKAT observations have revealed the propagation of radio pulses from pulsar A through the plasma-filled magnetosphere of pulsar B has an extraordinary impact on its detected polarization properties.
We detected significant variations in linear PA and observed the appearance of large amounts of circular polarization that varies in handedness throughout the partially-transparent phases of the eclipses.
Our spectropolarimetric modelling of several partial-transparency windows shown in Figure~\ref{fig:spectra} failed to recover a substantial frequency dependence nor a constrained generalised rotation measure. 
Hence the simplistic picture of Faraday conversion in a plasma threaded by a uniform magnetic field cannot explain the detected circular polarization.
Instead, the observed variations are all readily replicated by the radiative transfer model of \citet{Lyutikov2005b}.
Here, the pair-plasma filled magnetosphere of pulsar B acts as a polarizer in which radio waves with linear polarization states that are perpendicular to the local magnetic field of pulsar B will be preferentially absorbed.
As a result, any radiation that escapes the plasma should display a large degree of polarization that is aligned with the magnetic field direction.
However, differences in refractive index introduces a phase delay between the transmitted linear modes, which leads to a conversion of the initial linear polarization to circular polarization when the modes are coherently recombined after the transmitted wave exits the magnetosphere of pulsar B.
This is similar to a combination of Faraday-conversion-like effect and synchrotron-cyclotron absorption that was proposed to explain a flip in circular handedness detected in the black-widow pulsar PSR~B1744$-$24A \citep{Li2023}.
There, Faraday conversion is suggested to arise from a reversal in the line of sight magnetic field of the low-mass stellar companion, as opposed to birefringence within the hot pair-plasma surrounding a neutron star. 

\begin{figure}
    \centering
    \includegraphics[width=\linewidth]{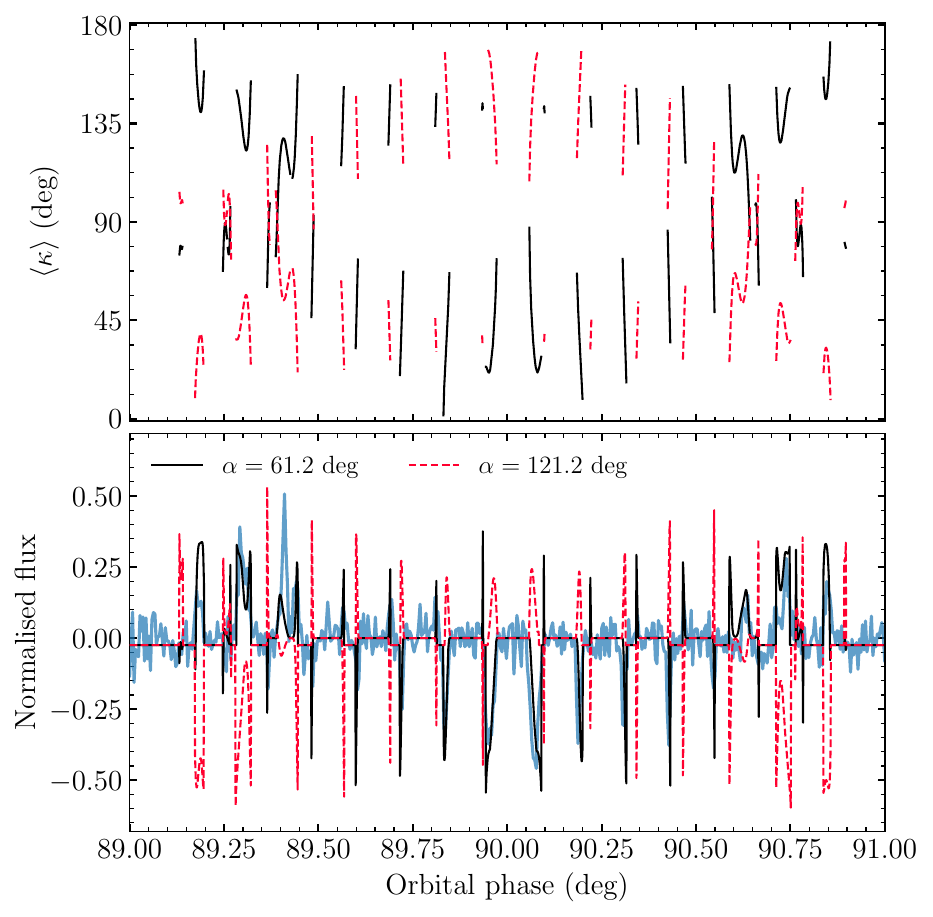}
    \caption{Stokes $V$ dependence on the line of sight magnetic field direction. Top panel shows the simulated direction of pulsar B's magnetic field averaged over our line of sight for two magnetic inclination angle ($\alpha$) values. Gaps indicate where $\tau_{\nu} = 0$ or $\tau_{\nu} \geq 1$. Bottom panel shows the computed Stokes $V$ light curves for both choices of $\alpha$ (solid black and dashed red) compared against the observed data in light blue.}
    \label{fig:bfield}
\end{figure}

The changes in circular handedness throughout the eclipse can be ascribed to differences in the average magnetic field direction along our line of sight, where differences in refractive index between the polarization modes depend on the angle between the wave propagation direction and the local magnetic field vector \citep{Lyutikov2005b}.
As a consequence of this behaviour, we are able to use the changing Stokes $V$ handedness to break a degeneracy in the magnetic inclination angle ($\alpha$) of pulsar B, i.e the angle between the magnetic moment of the neutron star and its rotational axis.
Our previous modelling of the secular evolution in the MeerKAT eclipses found a symmetry between $\alpha$ and the reference rotation phase of pulsar B at each epoch, where bi-modal posterior distributions were recovered for both parameters \citep{Lower2024}.
For $\alpha$ the posterior displayed two peaks at $\sim$60\,deg and $\sim$120\,deg.
While the total intensity light curves generated using these two solutions of $\alpha$ (alongside a half-rotation difference in pulsar B phase) are identical, the Stokes $V$ light curve is not.
We demonstrate this in Figure~\ref{fig:bfield}, where we plot the predicted average line-of-sight magnetic field direction of pulsar B, $\langle \kappa \rangle$, for the two unique solutions of $\alpha$ and both the detected and simulated Stokes $V$ light curves.
Comparing the light curves in the bottom panel, the solution in which $\alpha = 61.2$\,deg returns the same approximate handedness of Stokes $V$ as the data, whereas $\alpha = 121.2$\,deg returns the opposite handedness.
The larger value of $\alpha$ also massively over-predicts the induced Stokes $V$ amplitude in the ingress and egress phases, while under-predicting the amplitude at superior conjunction.
Hence, the $\alpha = 61.2$\,deg solution is correct.
Looking at the solid black curves in the top panel, we can see that LHC polarization is indeed detected whenever the magnetic field of pulsar B is pointed away from us ($\langle \kappa \rangle > 90$\,deg), switching to RHC when the magnetic field is pointed toward us ($\langle \kappa \rangle < 90$\,deg).
Combined with measurements from precision timing and scintillation studies of pulsar A \citep{Hu2022, Askew2024}, and total intensity light-curve modelling \citep{Lower2024}, we have been able to derive, for the first time, the precise magnetic and viewing geometry of a pulsar from radio observations.

The lack of significant RM changes throughout the eclipse envelope or at orbital phases immediately after the eclipses points to a low number density of mildly relativistic electrons within the magnetosheath and magnetotail of pulsar B.
Using the scatter in the UHF RM measurements, we can derive an upper-limit on the number density of electrons ($n_{e}$) from \citep{Yuen2012}
\begin{equation}
    \Delta{\rm RM} \approx 0.81 \langle n_{e} \rangle \Big(\frac{L}{1\,{\rm pc}}\Big) \Big(\frac{\langle B \rangle}{1\,{\rm \mu G}}\Big),
\end{equation}
where $L$ and $\langle B \rangle$ are the path length and average magnetic field strength in units of pc and $\mu G$ respectively.
Substituting in $\Delta{\rm RM} \lesssim 0.6$\,rad\,m$^{-2}$ alongside the approximate values of $L \sim 1.5 \times 10^{-8}$\,pc and $\langle B \rangle \sim 2$\,G assumed in \citet{Yuen2012} returns $n_{e} \lesssim 25$\,cm$^{-3}$.
A lack of detectable RM variations could also result from turbulence within the magnetotail, where changes in magnetic field direction along the line of sight result in a net-zero change in RM.
Multi-path propagation through this turbulent medium may result in a frequency-dependent reduction in detected linear polarization around superior conjunction (see \citealt{Beniamini2022}), yet we do not observe such an effect in the eclipse-averaged L-band and UHF light curves in Figure~\ref{fig:incoherent}.
Observations conducted at much lower observing frequencies with the Murchison Widefield Array or SKA-Low telescope may provide stronger constraints on such scattering-induced depolarization. 
Several previous works suggested that the linear PA of pulsar A would also have a magnetic-field direction dependence imparted by differential synchrotron absorption between the orthogonally polarized linear modes \citep{Rafikov2005, Lyutikov2005b, Yuen2012}.
Without seeing an obvious frequency dependence, this effect could be mistaken for Faraday rotation. 
All four example polarization light curves in Figure~\ref{fig:coherent} show clear variations in PA.
Our comparison to the simulated polarization properties in Figure~\ref{fig:comparison} show these variations are indeed the result of differential synchrotron absorption.
Hence, the changes in PA identified by \citet{Yuen2012} were therefore likely to have been a result of this effect as opposed to Faraday rotation.

The overall consistency between the observed and simulated polarization properties of pulsar A in Figure~\ref{fig:comparison} lends substantial support to the validity of the \citet{Lyutikov2005b} eclipse model.
From previous works, modelling of the total intensity light curve does provide some constraints on the general plasma and magnetic field properties.
Our resulting fits to the MeerKAT light curves in \citet{Lower2024} are consistent with a $R_{\rm mag} \sim 1.3 \times 10^{10}$\,cm and a pair multiplicity of $\lambda \sim 3.6 \times 10^{5}$.
Both are slightly smaller than previous estimates based on eclipse light curve modelling \citep{Lyutikov2005b, Breton2012}, however their overall conclusions around the plasma-content and origin of the large multiplicity remain unchanged.
The difference in magnetosphere size is likely a consequence of the model not taking into account changes in the elliptical orbital geometry due to periastron advance, which would alter the eclipse duration.
Non-axisymmetries in the magnetic topology of pulsar B imparted by the wind from pulsar A and secular changes due to spin-precession may also be contributing factors.
Our choice of magnetic field strength at the truncation radius has little to no impact on the simulated eclipse shape or polarization properties for values above $B_{\rm mag} \gtrsim 7$\,G.
Light curves generated using weaker $B_{\rm mag}$ values appear shallower and display additional transparency windows at half the rotation period of pulsar B around superior conjunction, neither of which are consistent with the data.
Hence, while it is remarkable that the \citet{Lyutikov2005b} model replicates much of the observed eclipse phenomenology, there are obvious improvements that need to be implemented in order to obtain an even clearer picture of the magnetosphere of pulsar B.
This includes adding in the aforementioned non-axisymmetries and light propagation effects, alongside a more refined treatment of particle transport within the double pulsar system informed via modern particle-in-cell simulations and theoretical models \citep{Zhong2024, Lyutikov2022a}.

\section{Summary and conclusions}\label{sec:conc}

We have presented the first complete spectropolarimetric analysis of the double pulsar eclipses using a set of calibrated MeerKAT observations, opening a new window to studying this incredible system.
Our results indicate the number density of mildly relativistic particles in the magnetotail of pulsar B is insufficient to alter the nominal Faraday rotation or linear PA of pulsar A.
Instead, the PA variations and large amounts of induced circular polarization that we observe qualitatively match the radiative transfer model of \citet{Lyutikov2005b}. 
This confirms that the eclipses arise from synchrotron absorption in the relativistic pair-plasma confined to the truncated magnetosphere of pulsar B, independent to the success of their total intensity model.
We also showed the changing circular handedness is linked to average line of sight magnetic field direction of pulsar B, allowing us to break a symmetry between the magnetic inclination angle and per-epoch rotation phase of the pulsar.
Combined with previous timing and polarization studies of pulsar A \citep{Kramer2021a, Kramer2021b, Hu2022}, modelling of the eclipses in total intensity \citep{Lower2024}, and upcoming scintillation results \citep{Askew2024}, we now know the full three-dimensional orbital, viewing and magnetic geometry of pulsar B.

While the consistency between our polarization observations and the theoretical model is remarkable, there are some key limitations that are yet to be overcome.
The light curve model does not fully account for the differences in the light propagation time from pulsar A to pulsar B due to periastron precession over time, nor the relativistic light-bending effect \citep{Hu2022}.
Additionally, the assumed perfectly axisymmetric dipole magnetic geometry is known to not be an accurate representation of the true structure of pulsar B's magnetosphere.
Numerical simulations have shown that the windward side of the magnetosphere will be more strongly compressed by the relativistic wind from pulsar A than the leeward side, imparting a substantial asymmetry to the overall shape (e.g, \citealt{Arons2005,  Zhong2024}). 
Inclusion of such effects in the model could help overcome noted deviations towards the eclipse edges.
The development of enhanced eclipse models that take these effects into account are additionally motivated by both near-future enhancements to the sensitivity of MeerKAT through the `MeerKAT+' extension project and its eventual integration into the SKA-Mid telescope.
These will enable even higher time and frequency resolution studies of the birefringent nature of the magnetospheric plasma and propagation within the magnetotail of pulsar B than are possible with current generation radio telescopes.
These future data sets, combined with advanced models, will ensure the double pulsar remains one of the best astrophysical laboratories for testing our theories of fundamental physics.

\section*{Acknowledgements}
We thank Maxim Lyutikov, Don Melrose and Kaustubh Rajwade for insightful discussions on polarized radiative transfer, and the anonymous referee for their comments on the manuscript.
The MeerKAT telescope is operated by the South African Radio Astronomy Observatory (SARAO), which is a facility of the National Research Foundation, an agency of the Department of Science and Innovation. 
SARAO acknowledges the ongoing advice and calibration of GPS systems by the National Metrology Institute of South Africa (NMISA) and the time space reference systems department department of the Paris Observatory. 
PTUSE was developed with support from the Australian SKA Office and Swinburne University of Technology. 
This work made use of the OzSTAR national HPC facility at Swinburne University of Technology.
MeerTime data is housed on the OzSTAR supercomputer.
The OzSTAR program receives funding in part from the Astronomy National Collaborative Research Infrastructure Strategy (NCRIS) allocation provided by the Australian Government.
MEL and LSO acknowledge support from the Royal Society International Exchange grant IES\textbackslash R1\textbackslash 231332.
MK acknowledges significant support from the Max-Planck Society (MPG) and the MPIfR contribution to the PTUSE hardware. 
RMS acknowledges support through Australian Research Council (ARC) Future Fellowship FT190100155. 
RPB acknowledges support from the European Research Council (ERC) under the European Union’s Horizon 2020 research and innovation programme (grant agreement no. 715051; Spiders).
MB is supported by ARC CE230100016.
Part of the work was undertaken as part of the ARC Centre of Excellence for Gravitational Wave Discovery (OzGrav; CE170100004 and CE230100016).
LSO acknowledges support from Magdalen College, Oxford.
VVK acknowledges financial support from the European Research Council (ERC) starting grant "COMPACT" (Grant agreement number 101078094). 
This work has made use of NASA's Astrophysics Data System.

\section*{Data Availability}

The data and data products used in this work are available upon reasonable request to the corresponding author.



\bibliographystyle{mnras}
\bibliography{main} 




\bsp	
\label{lastpage}
\end{document}